\documentclass{aastex}
\usepackage{graphicx}
\usepackage{times}
\begin{document}


\title{
BVRI Photometry of SN 2013ej in M74
}

\author{Michael W. Richmond}
\shorttitle{SN 2013ej in M74}
\shortauthors{M. W. Richmond }
\affil{
School of Physics and Astronomy, Rochester Institute of Technology,
84 Lomb Memorial Drive, Rochester, NY, 14623 USA
}
\email{mwrsps@rit.edu}


\begin{abstract}

I present BVRI photometry of the type IIP supernova 2013ej in M74
from $1$ to $179$ days after its discovery.
These photometric measurements and spectroscopic data from
the literature are combined via the expanding photosphere method
to estimate the distance to the event,
which is consistent with that derived by other methods.
After correcting for extinction
and adopting a distance modulus of 
$(m - M) = 29.80$ mag to M74,
I derive absolute magnitudes 
$M_B = -17.36$,
$M_V = -17.47$,
$M_R = -17.64$
and 
$M_I = -17.71$.
The differences between visual measurements and CCD $V$-band measurements
of SN 2013ej are similar to those determined for type Ia supernovae
and ordinary stars.

\end{abstract}


\section{Introduction}

On UT 2013 July 25.45, the Lick Observatory Supernova Search (LOSS)
detected a new point source in the nearby galaxy M74 (NGC 628);
when the object appeared again and brighter the next night, 
LOSS alerted other astronomers to the presence of this new object.
Within days, spectroscopy revealed it to be a young type II supernova,
designated SN 2013ej
\citep{Kim2013}.
Because its host is so nearby (less than 10 Mpc; see section 5)
and so well studied,
and because the event was caught within a few days of
the explosion, 
SN 2013ej provides a fine opportunity for us to study
the properties of a massive star before and after it
undergoes core collapse.

I present here photometry of SN 2013ej in the 
$BVRI$ passbands obtained at the RIT Observatory,
starting one day after the announcement and
continuing for a span of 179 days.
Section 2 describes the observational procedures,
the reduction of the raw images,
and the methods used to extract instrumental magnitudes.
In section 3, I explain how the instrumental
quantities were transformed to the standard
Johnson-Cousins magnitude scale.
I illustrate the light curves and color curves
of SN 2013ej in section 4 and
comment briefly on their properties.
In section 5,
I discuss extinction along the line of sight to this event.
In section 6, 
I discuss attempts to measure the distance
to M74,
and use the Expanding Photosphere Method (EPM)
to perform my own estimate;
I adopt a distance and convert the apparent 
magnitudes at peak to absolute magnitudes.
Visual measurements of this event collected by the
American Association of Variable Star Observers 
(AAVSO) are compared to CCD $V$-band measurements in
section 7.
I summarize the results of this study in section 8.

\section{Observations}

This paper contains measurements made at the RIT Observatory,
near Rochester, New York.
The RIT Observatory is located on the campus of the 
Rochester Institute of Technology, at 
longitude 77:39:53 West, latitude +43:04:33 North,
and an elevation of 168 meters above sea level.
The eastern horizon is bright and dominated by a
large pine tree.  
Measurements during the first two weeks, and particularly
on the very first night, were taken at low airmass and 
not far from the tree's branches.
I used a Meade LX200 f/10 30-cm telescope
and SBIG ST-8E camera, which features a Kodak KAF1600 CCD chip
and astronomical filters made to the 
\citet{Bess1990} prescription;
with $3 \times 3$ binning,
the plate scale is 
$1{\rlap.}^{''}85$
per pixel.
To measure SN 2013ej, 
I took a series of 30-second exposures through
each filter, using the autoguider if possible;
the only guide star was very faint in the $B$-band,
so most of those images were unguided.
The number of exposures 
per filter ranged from 10, at early times, to 15 or 30 at late times.
I typically discarded a few images in each series
due to trailing.
I acquired dark and flatfield images each night,
except for UT Dec 17;
the images from that night were reduced using dome flats
taken the following evening.
In most cases, I chose to use dome flats over twilight sky flatfield images.

I combined 10 dark images each night to create a master
dark frame, and 10 flatfield images in each filter to create
a master flatfield frame.
After applying the master dark and flatfield images in the
usual manner,
I examined each cleaned target image by eye.
I discarded trailed and blurry images and
measured the Full Width at Half Maximum (FWHM) of those remaining.
The XVista 
\citep{Tref1989}
routines 
{\tt stars} 
and
{\tt phot}
were used to find stars 
and to extract their instrumental magnitudes,
respectively,
using a synthetic aperture with radius
of 4 pixels (= $7{\rlap.}^{''}4$),
slightly larger than the FWHM (which was typically 
$4''$ to $5''$).

\begin{figure}
 \plotone{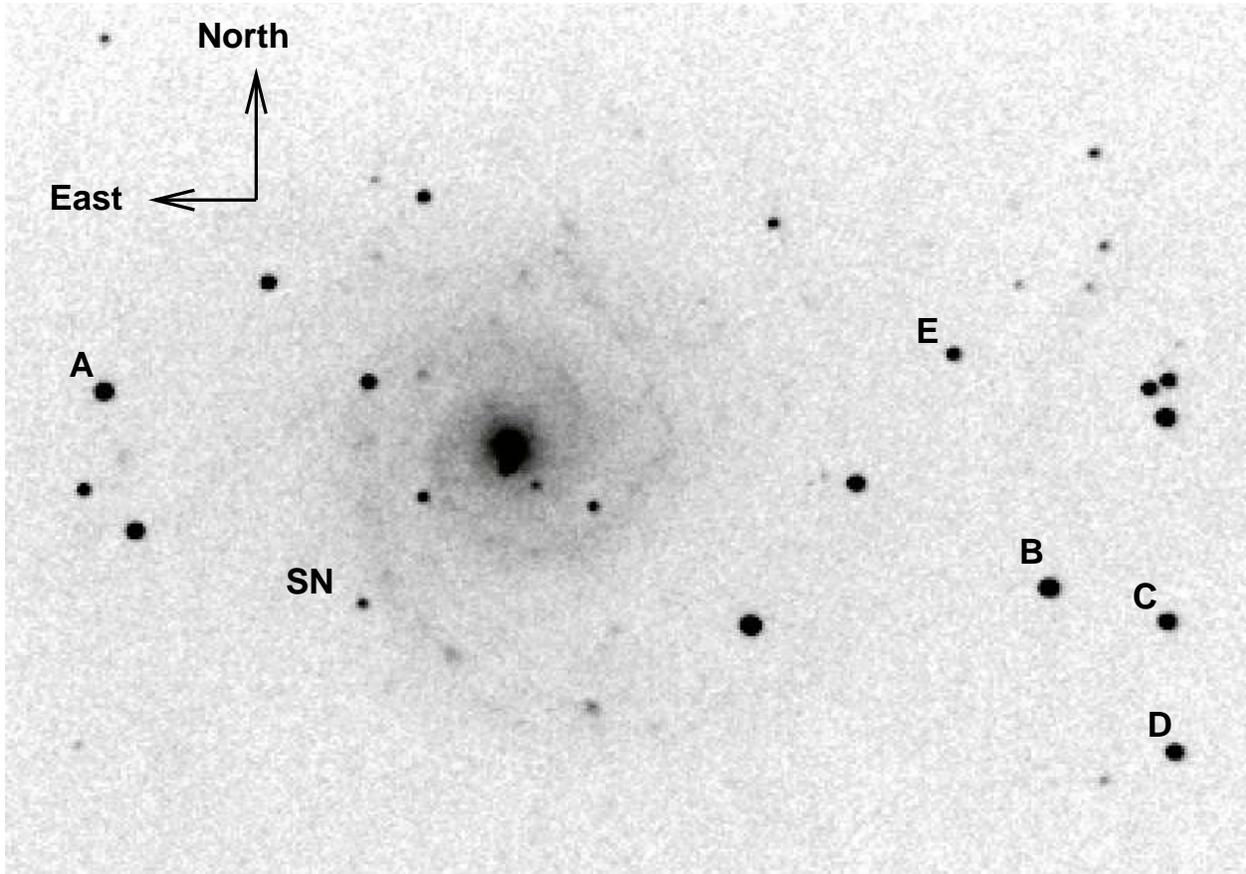}
 \caption{An R-band image of M74 from RIT,
          15 x 30 seconds exposure time,
          showing stars used to calibrate measurements 
          of SN 2013ej.  North is up, East to the left.
          The field of view is roughly 12 by 9 arcminutes.
          \label{fig:chartlabel} }
\end{figure}

As
Figure \ref{fig:chartlabel}
shows,
SN 2013ej lies in the outskirts of
one of the spiral arms of M74.
How much light from other objects in the area
falls into the aperture used to measure
the supernova?
I examined high-resolution HST images of
this region,
using ACS WFC data in the F435W, F555W and F814W filters
originally taken as part of proposal GO-10402
(PI: Chandar).
See
\citet{Fras2014}
for a detailed analysis of the progenitor's
light in these images.
Within a circle of radius 
$7{\rlap.}^{''}4$
centered on the SN's position
are ten or so point sources of
roughly equal brightness,
with magnitudes of roughly $B \sim 25$, 
$I \sim 23$.
The combined light of these sources is too small to 
make a significant addition to the light of the
SN itself.
However, 
a considerably brighter source lies at
RA 01:36:48.55, Dec +15:45:26.5,
a distance of 
$7{\rlap.}^{''}7$ to the southeast of SN 2013ej.
Comparing it to the progenitor in the HST images,
I measure magnitudes of $B = 22.64$, $V = 21.15$, $I = 18.10$.
The $I$-band value agrees well with an entry in
the USNO B1.0 catalog 
\citep{Mone2003}.
Since this star lies at the edge of the synthetic aperture
used to measure the SN,
some of its light was attributed to the SN in my measurements.
In the $B$ and $V$ images, SN 2013ej was at least 3.9 magnitudes
brighter than this star at all times,
and so the contaminating flux was at most a few percent.
In the $R$ and $I$ images,
on the other hand, this star's light may have been important 
at late times.
In the last $I$-band measurement, for example,
roughly one-sixth of the measured light may have
come from this star.
Since the exact amount of contamination depends on
details of the seeing and shape of the point-spread function
on each night, 
I have made no correction for this effect;
but the late-time measurements reported here are slightly
brighter than they ought to be, especially in the red passbands.

\begin{figure}
 \plotone{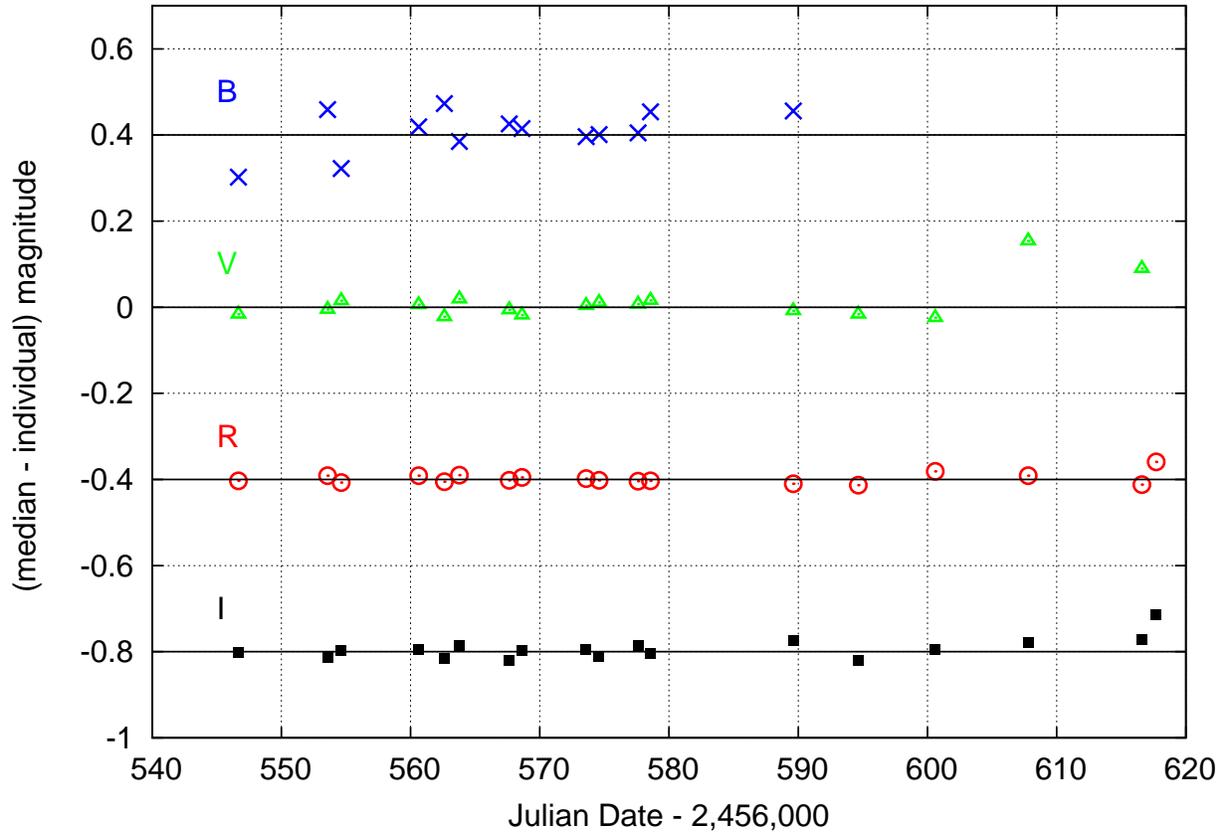}
 \caption{Difference between instrumental magnitudes
          extracted from median-combined images
          and from individual images at RIT. 
          The values have been shifted for clarity
          by 0.4, 0.0, -0.4, -0.8 mag in 
          B, V, R, I, respectively.
          \label{fig:indivmedian} }
\end{figure}

Between July and early September, 2013, 
I measured instrumental magnitudes from each exposure
and applied inhomogeneous ensemble photometry
\citep{Hone1992}
to determine a mean value in each passband.
Starting on UT Sep 11, 2013,
in order to improve the signal-to-noise ratio,
I combined the good images for each passband
using a pixel-by-pixel median procedure
to yield a single image with lower noise levels.
I then extracted instrumental magnitudes
from this image in the manner described above.
In order to verify that this change in procedure
did not cause any systematic shift in the results,
I also measured magnitudes from the individual exposures
at these late times,
reduced them using ensemble photometry,
and compared the results to those measured from the 
median-combined images.
As Figure \ref{fig:indivmedian} shows,
there were no significant systematic differences.

\section{Photometric calibration}

In order to transform the instrumental measurements
into magnitudes in the standard Johnson-Cousins BVRI
system,
I used a set of local comparison stars,
supplied by the AAVSO
in their chart 12459CA.
These reference stars are listed in Table
\ref{tab:compstars},
and
Figure \ref{fig:chartlabel}
shows their location.

\begin{center}
 \begin{table*}[ht]
   \caption{Photometry of comparison stars}
   \label{tab:compstars}
   {\small
    \hfill{}
    \begin{tabular}{l l l c c c c}
    \hline
    Star & RA (J2000) &  Dec (J2000) &  B  & V & R & I  \\
    \hline

    A  &  01:36:58.63 & +15:47:46.7 & $13.012 \pm 0.019$ & $12.510 \pm 0.019$ & $12.154 \pm 0.019$ & $11.834 \pm 0.019$ \\
    B  &  01:36:19.55 & +15:45:22.4 & $13.848 \pm 0.026$ & $13.065 \pm 0.022$ & $12.622 \pm 0.025$ & $12.152 \pm 0.027$ \\
    C  &  01:36:14.64 & +15:44:58.6 & $14.338 \pm 0.029$ & $13.692 \pm 0.024$ & $13.329 \pm 0.029$ & $12.964 \pm 0.030$ \\
    D  &  01:36:14.60 & +15:43:39.5 & $14.832 \pm 0.027$ & $13.912 \pm 0.023$ & $13.416 \pm 0.026$ & $12.939 \pm 0.030$ \\
    E  &  01:36:23.06 & +15:47:45.4 & $15.192 \pm 0.034$ & $14.613 \pm 0.027$ & $14.275 \pm 0.034$ & $13.915 \pm 0.036$ \\

    \hline
    \hline
  \end{tabular}
  }
 \end{table*}
\end{center}

In order to 
correct for differences between the RIT equipment and the
Johnson-Cousins system,
I observed the standard fields
PG1633+009 
and
PG2213-006
\citep{Land1992}
on several nights
and compared the instrumental magnitudes to catalog values.
Linear fits to the differences as a function of color
yielded the following relationships:
\begin{eqnarray}
  B  &=  b  +  (0.231 \pm 0.012) * (b - v)  +  Z_B   \\
  V  &=  v  -  (0.079 \pm 0.017) * (v - r)  +  Z_V   \\
  R  &=  r  -  (0.087 \pm 0.021) * (r - i)  +  Z_R   \\
  I  &=  i  -  (0.018 \pm 0.040) * (r - i)  +  Z_I   
\end{eqnarray}
In the equations above, 
lower-case symbols represent instrumental magnitudes,
upper-case symbols Johnson-Cousins magnitudes,
and $Z$ the zeropoint in each band.
Stars A, B, C, D and E were used to set the 
zeropoint for each image.
Table \ref{tab:ritphot} lists our calibrated measurements 
of SN 2013ej made at RIT.
The first column shows the mean Julian Date of all the exposures
taken during each night.
In most cases, the span between the first and last exposures was less than 
$0.04$ days, but on a few nights, clouds interrupted the sequence of 
observations.  
Contact the author for a dataset providing the Julian Date 
of each measurement individually.

\begin{center}
 \begin{deluxetable}{l l l l l l}
   \tablecaption{Photometry of SN 2013ej \label{tab:ritphot}}
   \tablehead{
     \colhead{JD-2456500} &
     \colhead{B} &
     \colhead{V} &
     \colhead{R} &
     \colhead{I} &
     \colhead{comments}
   }

\startdata

     0.71  &  $12.945 \pm 0.059$  &  $12.999 \pm 0.025$  &  $12.972 \pm 0.060$  &  $12.967 \pm 0.056$  &   high airmass  \\  
     3.80  &  $12.714 \pm 0.035$  &  $12.647 \pm 0.012$  &  $12.566 \pm 0.021$  &  $12.537 \pm 0.025$  &   cirrus  \\  
     4.73  &  $12.693 \pm 0.020$  &  $12.615 \pm 0.019$  &  $12.509 \pm 0.027$  &  $12.446 \pm 0.058$  &   cirrus  \\  
     6.81  &  $12.624 \pm 0.047$  &  $12.524 \pm 0.021$  &  $12.404 \pm 0.018$  &  $12.373 \pm 0.025$  &   \\  
     8.75  &  $12.668 \pm 0.048$  &  $12.522 \pm 0.014$  &  $12.370 \pm 0.030$  &  $12.275 \pm 0.037$  &   clouds  \\  
     9.73  &  $12.700 \pm 0.056$  &  $12.513 \pm 0.042$  &  $12.350 \pm 0.034$  &  $12.318 \pm 0.041$  &   \\  
    10.83  &  $12.715 \pm 0.028$  &  $12.553 \pm 0.032$  &  $12.321 \pm 0.043$  &  $12.291 \pm 0.052$  &   clouds  \\  
    14.69  &  $12.964 \pm 0.056$  &  $12.527 \pm 0.075$  &  $12.297 \pm 0.051$  &  $12.239 \pm 0.034$  &   clouds  \\  
    15.70  &  $12.973 \pm 0.036$  &  $12.548 \pm 0.013$  &  $12.303 \pm 0.015$  &  $12.219 \pm 0.030$  &   \\  
    19.70  &  $13.239 \pm 0.032$  &  $12.586 \pm 0.026$  &  $12.310 \pm 0.028$  &  $12.176 \pm 0.035$  &   \\  
    20.68  &  $13.351 \pm 0.086$  &  $12.601 \pm 0.028$  &  $12.309 \pm 0.043$  &  $12.149 \pm 0.042$  &   cirrus  \\  
    21.70  &  $13.421 \pm 0.084$  &  $12.651 \pm 0.032$  &  $12.339 \pm 0.031$  &  $12.177 \pm 0.043$  &   \\  
    24.69  &  $13.564 \pm 0.094$  &  $12.748 \pm 0.039$  &  $12.378 \pm 0.028$  &  $12.237 \pm 0.051$  &   \\  
    25.69  &  $13.734 \pm 0.137$  &  $12.787 \pm 0.058$  &  $12.429 \pm 0.026$  &  $12.211 \pm 0.034$  &   nearby moon  \\  
    28.70  &  $13.831 \pm 0.056$  &  $12.864 \pm 0.038$  &  $12.470 \pm 0.025$  &  $12.255 \pm 0.036$  &   \\  
    29.66  &  $13.939 \pm 0.090$  &  $12.904 \pm 0.027$  &  $12.507 \pm 0.030$  &  $12.290 \pm 0.038$  &   \\  
    33.68  &  $14.109 \pm 0.085$  &  $13.026 \pm 0.045$  &  $12.574 \pm 0.023$  &  $12.340 \pm 0.028$  &   \\  
    38.62  &  $14.346 \pm 0.140$  &  $13.142 \pm 0.034$  &  $12.677 \pm 0.021$  &  $12.403 \pm 0.062$  &   clouds  \\  
    39.79  &  $14.406 \pm 0.039$  &  $13.164 \pm 0.026$  &  $12.686 \pm 0.019$  &  $12.446 \pm 0.030$  &   \\  
    41.75  &  $14.442 \pm 0.052$  &  $13.228 \pm 0.024$  &  $12.734 \pm 0.020$  &  $12.491 \pm 0.045$  &   clouds  \\  
    44.62  &  $14.495 \pm 0.096$  &  $13.291 \pm 0.037$  &  $12.765 \pm 0.024$  &  $12.514 \pm 0.035$  &   high airmass  \\  
    46.68  &  $14.642 \pm 0.094$  &  $13.302 \pm 0.028$  &  $12.824 \pm 0.027$  &  $12.542 \pm 0.051$  &   \\  
    53.57  &  $14.747 \pm 0.084$  &  $13.438 \pm 0.031$  &  $12.902 \pm 0.026$  &  $12.587 \pm 0.029$  &   nearby moon  \\  
    54.63  &  $14.889 \pm 0.048$  &  $13.450 \pm 0.039$  &  $12.909 \pm 0.016$  &  $12.627 \pm 0.021$  &   nearby moon  \\  
    60.62  &  $14.932 \pm 0.055$  &  $13.538 \pm 0.055$  &  $13.007 \pm 0.018$  &  $12.699 \pm 0.026$  &   \\  
    62.62  &  $14.993 \pm 0.056$  &  $13.570 \pm 0.041$  &  $13.006 \pm 0.016$  &  $12.691 \pm 0.026$  &   clouds  \\  
    63.77  &  $14.955 \pm 0.072$  &  $13.618 \pm 0.025$  &  $13.037 \pm 0.020$  &  $12.746 \pm 0.025$  &   \\  
    67.62  &  $15.082 \pm 0.070$  &  $13.634 \pm 0.041$  &  $13.080 \pm 0.020$  &  $12.759 \pm 0.022$  &   \\  
    68.62  &  $15.119 \pm 0.060$  &  $13.650 \pm 0.046$  &  $13.110 \pm 0.022$  &  $12.780 \pm 0.030$  &   hazy  \\  
    73.59  &  $15.285 \pm 0.067$  &  $13.733 \pm 0.031$  &  $13.172 \pm 0.037$  &  $12.864 \pm 0.031$  &   \\  
    74.59  &  $15.234 \pm 0.068$  &  $13.771 \pm 0.040$  &  $13.191 \pm 0.021$  &  $12.900 \pm 0.024$  &   \\  
    77.60  &  $15.321 \pm 0.065$  &  $13.858 \pm 0.038$  &  $13.230 \pm 0.022$  &  $12.947 \pm 0.032$  &   clouds  \\  
    78.56  &  $15.357 \pm 0.093$  &  $13.868 \pm 0.043$  &  $13.280 \pm 0.021$  &  $12.967 \pm 0.027$  &   \\  
    89.58  &  $15.807 \pm 0.077$  &  $14.211 \pm 0.034$  &  $13.599 \pm 0.030$  &  $13.300 \pm 0.037$  &   clouds  \\  
    94.65  &  \nodata             &  $14.299 \pm 0.043$  &  $13.913 \pm 0.024$  &  $13.602 \pm 0.039$  &   \\  
    96.62  &  $16.412 \pm 0.149$  &  $14.907 \pm 0.051$  &  $14.174 \pm 0.035$  &  $13.811 \pm 0.050$  &   \\  
   100.58  &  $17.166 \pm 0.161$  &  $15.776 \pm 0.079$  &  $14.954 \pm 0.047$  &  $14.631 \pm 0.066$  &   clouds  \\  
   107.79  &  $18.193 \pm 0.311$  &  $16.612 \pm 0.137$  &  $15.388 \pm 0.062$  &  $15.044 \pm 0.082$  &   \\  
   110.64  &  \nodata             &  $16.309 \pm 0.133$  &  $15.378 \pm 0.082$  &  $15.100 \pm 0.119$  &   clouds  \\  
   113.64  &  \nodata             &  $16.572 \pm 0.160$  &  $15.584 \pm 0.083$  &  $15.219 \pm 0.110$  &   clouds  \\  
   116.59  &  \nodata             &  $16.478 \pm 0.123$  &  $15.501 \pm 0.064$  &  $15.180 \pm 0.090$  &   \\  
   117.71  &  \nodata             &  $16.444 \pm 0.135$  &  $15.497 \pm 0.078$  &  $15.132 \pm 0.107$  &   cirrus  \\  
   130.48  &  \nodata             &  $16.889 \pm 0.247$  &  $15.763 \pm 0.110$  &  $15.674 \pm 0.175$  &   cirrus  \\  
   143.52  &  \nodata             &  $17.235 \pm 0.237$  &  $16.001 \pm 0.100$  &  $15.561 \pm 0.149$  &   clouds  \\  
   155.49  &  \nodata             &  $17.089 \pm 0.217$  &  $16.144 \pm 0.109$  &  $16.132 \pm 0.196$  &   \\  

    \enddata
 \end{deluxetable}
\end{center}

The uncertainties listed in Table
\ref{tab:ritphot}
incorporate the uncertainties in instrumental magnitudes
and in the offset used to shift the instrumental values
to the standard scale, added in quadrature.
As a check on their size,
I chose a region of the light curve,
$40 < {\rm JD} - 2456500 < 80$,
in which the magnitude appeared to be a linear
function of time.
I fit
a straight line to the measurements in each passband,
weighting each point based on its uncertainty;
the results are shown in 
Table \ref{tab:linear}.
The reduced $\chi^2$ values are all
less than $1.0$, 
which 
suggests that the tabulated uncertainties 
slightly overestimate 
the random scatter from one measurement to the next.

\begin{center}
 \begin{deluxetable}{l c c }
   \tablecaption{Linear fit to light curves $40 < {\rm JD - 2456500} < 80$  \label{tab:linear} }
   \tablehead{
     \colhead{Passband} &
     \colhead{slope (mag/day)} &
     \colhead{reduced $\chi^2$}
   }

   \startdata

    B  &  $0.0238 \pm 0.0012$  &   0.6 \\
    V  &  $0.0167 \pm 0.0004$  &   0.3 \\
    R  &  $0.0141 \pm 0.0003$  &   0.5 \\
    I  &  $0.0131 \pm 0.0006$  &   0.8 \\

    \enddata
 \end{deluxetable}
\end{center}

\section{Light curves}

\begin{figure}
 \plotone{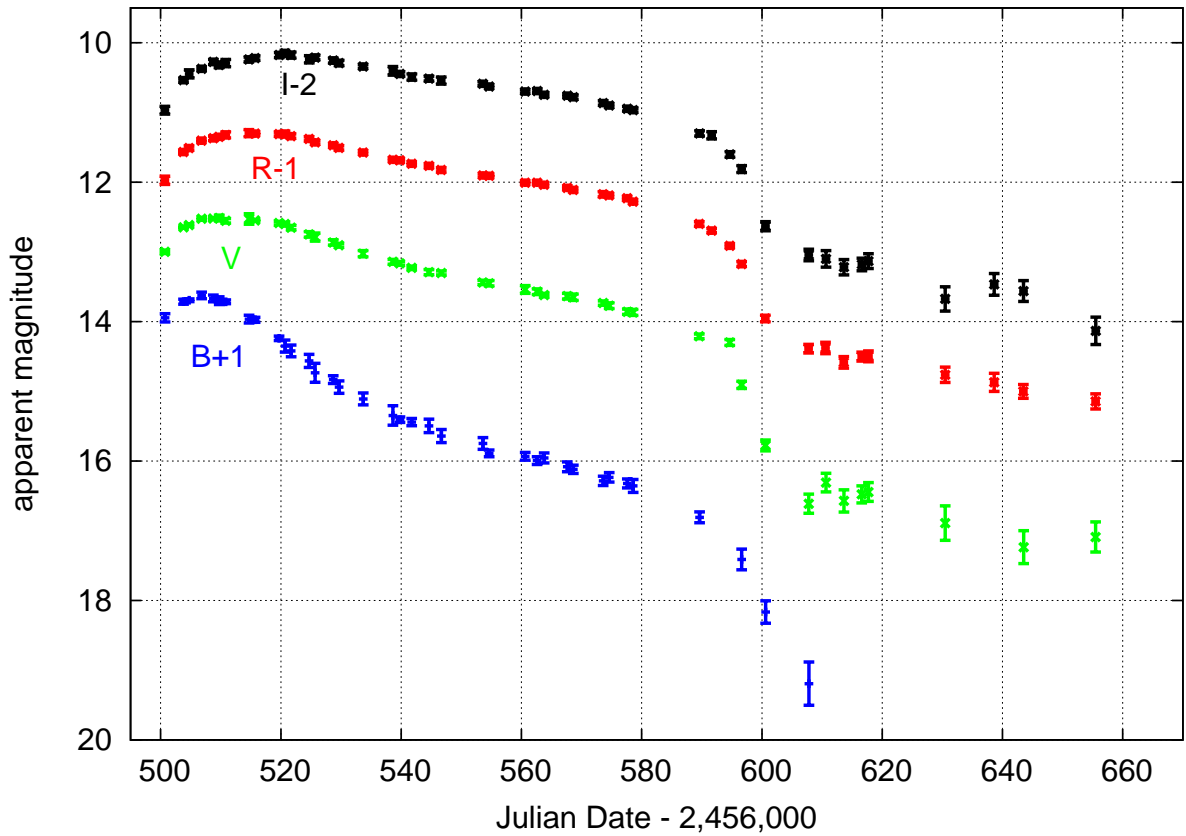}
 \caption{Light curves of SN 2013ej measured at RIT Observatory.
          The $B$, $R$ and $I$ data have been offset vertically
          for clarity.
          No correction for extinction has been made.
          \label{fig:lightcurves} }
\end{figure}

The light curves in each passband,
uncorrected for any extinction,
are shown in Figure \ref{fig:lightcurves}.
SN 2013ej is clearly a type IIP event,
defined by a period of roughly 60 days
during which the apparent brightness decreases
very slowly.
The plateau phase ends at 
Julian Date $\sim$ 2456590,
after which there is a sharp drop lasting a week or so.
The light curve then decreases at a moderate pace
for another month, to the end of the observations.

In order to determine the time and magnitude 
at peak light,
I fit second- and third-order polynomials to 
a subset of measurements around maximum light
in each passband.
Table \ref{tab:appmax} 
lists the results.
Maximum light occurs earliest in the
$B$-band 
and successively later at longer wavelengths.

\begin{center}
 \begin{deluxetable}{l r r }
   \tablecaption{Apparent magnitudes at maximum light \label{tab:appmax} }
   \tablehead{
     \colhead{Passband} &
     \colhead{JD-2456500} &
     \colhead{mag}
   }

   \startdata

    B        &  $\phantom{1.}7.3    \pm 0.2$  &   $12.64 \pm 0.01$ \\
    V        &  $12.1   \pm 1.0$  &   $12.48 \pm 0.02$ \\
    R        &  $14.9   \pm 1.0$  &   $12.28 \pm 0.01$ \\
    I        &  $19.0   \pm 2.0$  &   $12.17 \pm 0.02$ \\

    \enddata
 \end{deluxetable}
\end{center}

\begin{figure}
 \plotone{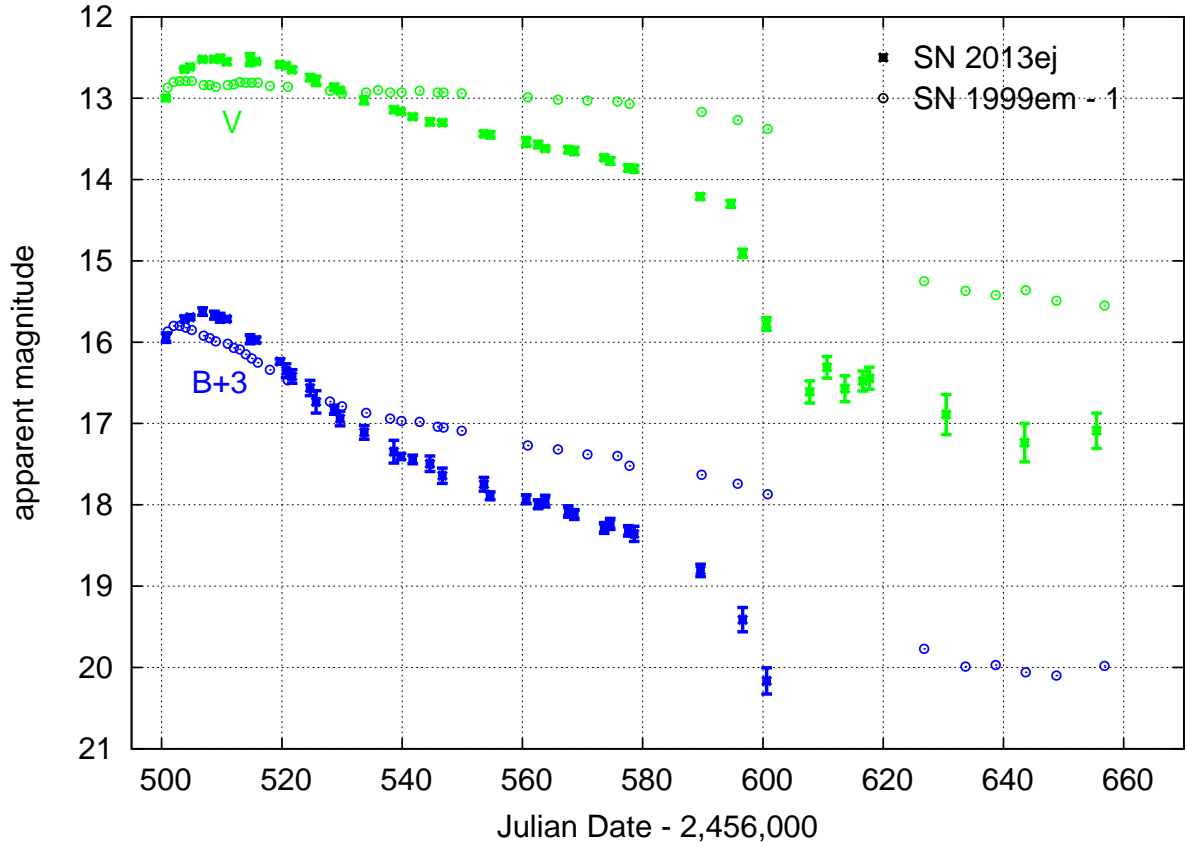}
 \caption{Light curves of SNe 2013ej and 1999em compared in
          the $B$ and $V$ passbands.  
          The measurements of SN 1999em have been shifted horizontally
          (by 5019 days) and vertically (by -1 mag) for easier
          comparison.
          \label{fig:1999em_bv} }
\end{figure}

\begin{figure}
 \plotone{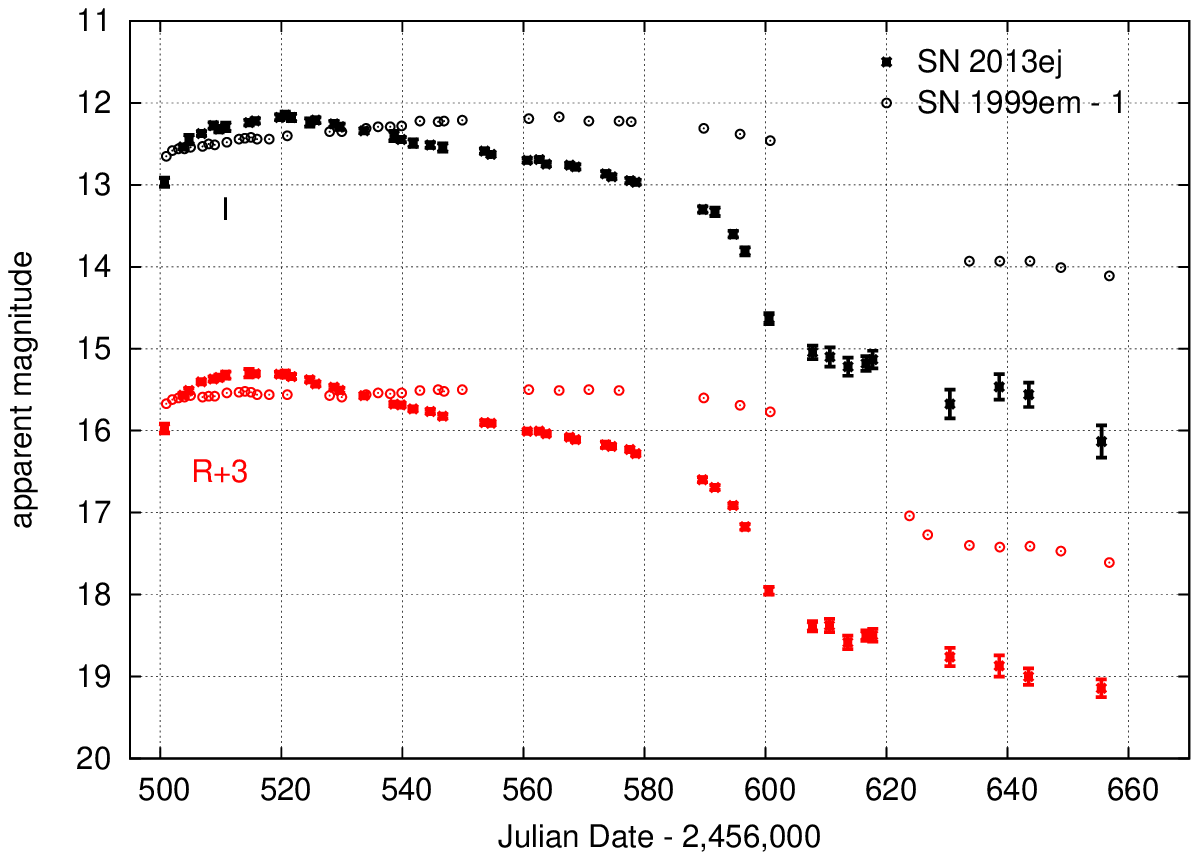}
 \caption{Light curves of SNe 2013ej and 1999em compared in
          the $R$ and $I$ passbands.  
          The measurements of SN 1999em have been shifted horizontally
          (by 5019 days) and vertically (by -1 mag) for easier
          comparison.
          \label{fig:1999em_ri} }
\end{figure}

The well-observed type IIP SN 1999em 
\citep{Leon2002}
provides a good 
comparison to SN 2013ej.  
In Figures 
\ref{fig:1999em_bv}
and
\ref{fig:1999em_ri},
one can see that SN 2013ej rises to and falls from an
early peak in all four passbands,
while SN 1999em has such a peak only in $B$;
its light curve is nearly flat in the other passbands.
The plateau phase ends slightly later in SN 1999em,
and the drop to the late-time decline is very similar.

\begin{figure}
 \plotone{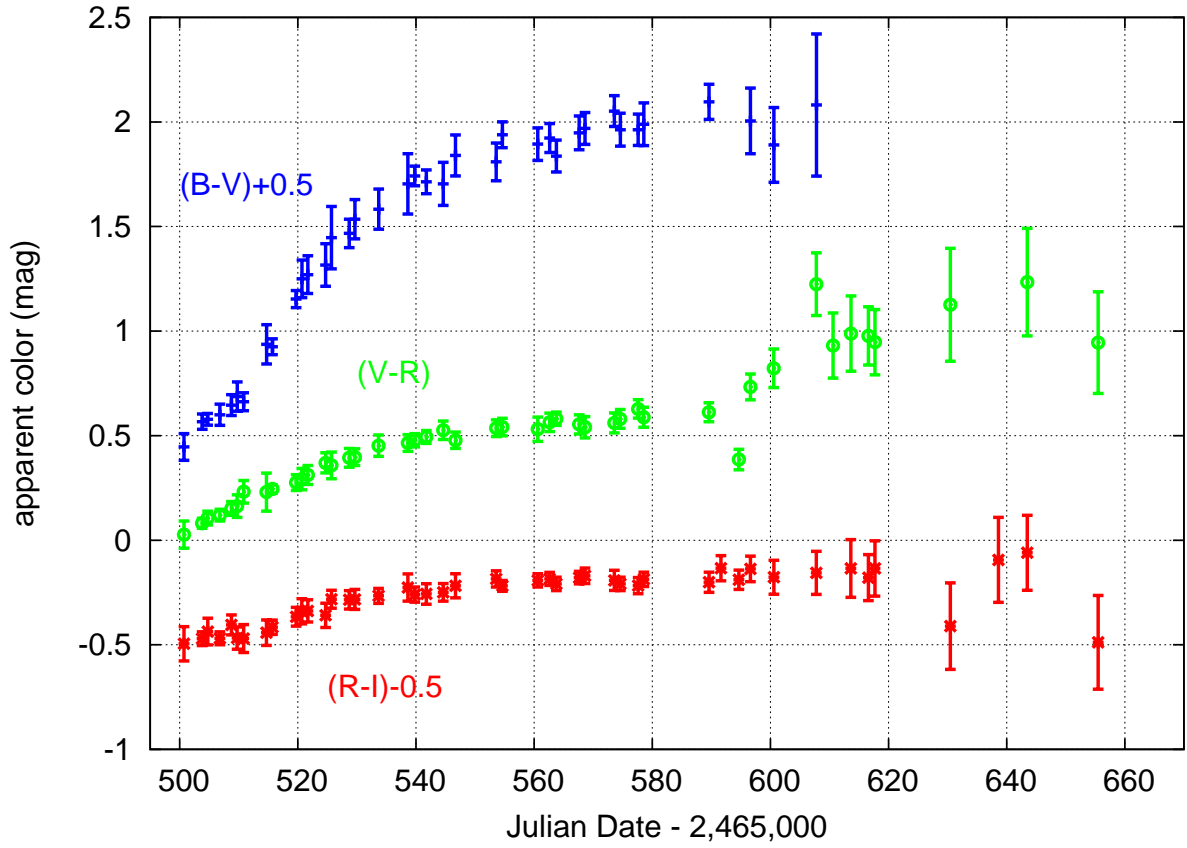}
 \caption{Color curves of SN 2013ej measured at RIT Observatory.
          The $(B-V)$ and $(R-I)$ data have been offset vertically
          for clarity.
          No correction for extinction has been made.
          \label{fig:colorcurves} }
\end{figure}

The colors of SN 2013ej changed considerably at the blue
end of the visible spectrum, but very little at the red end.
As Figure \ref{fig:colorcurves} indicates,
the $(B-V)$ color increased monotonically by
about 1.5 magnitudes over one hundred days.
The most rapid change occurred as the light curve
fell after maximum in $B$, but 
the increase then slowed during the plateau phase.
The $(R-I)$ color, on the other hand, remained
nearly constant, increasing by only 0.3 mag from
maximum light to the plateau phase.
The magnitude measurements after the end of the
plateau phase are so noisy that it is hard to see
any significant change in color at that time.

\begin{figure}
 \plotone{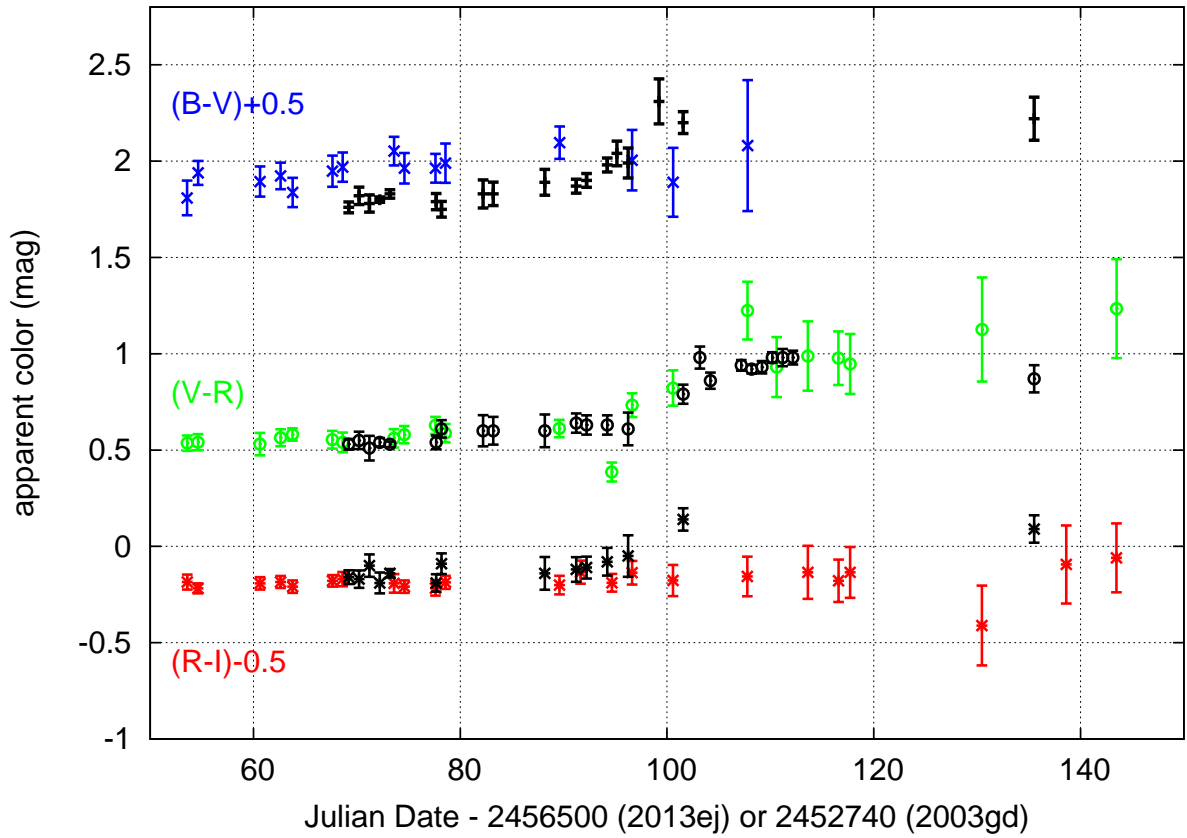}
 \caption{Color curves of SN 2013ej (colored symbols) compared with those of 
          SN 2003gd (black symbols).  
          The $(B-V)$ and $(R-I)$ data have been offset vertically
          for clarity.
          No correction for extinction has been made.
          \label{fig:comparecolors} }
\end{figure}

One can compare the colors of SN 2013ej to those of SN 2003gd, 
another type IIP SN in M74;
this will inform the discussion of extinction in section 5.
However, since SN 2003gd was discovered long after 
maximum light, 
this comparison is restricted largely to the plateau
phase, and one cannot align the two events in time
with any precision.
Figure \ref{fig:comparecolors} shows
the two events were very similar:
SN 2003gd had a slightly smaller $(B-V)$ color,
but only by 0.15 mag at most.

\section{Extinction}

There are several different methods one can use to 
estimate the extinction along the line of sight
to SN 2013ej.  
One can begin with the effects of dust and gas
within our own galaxy:
the foreground Milky Way reddening to M74 is
estimated by
\citet{Schl2011}
to be $E(B-V) = 0.062$.
Note that this value is an average based on
infrared maps with a beam size of order 6 arcminutes,
which subtends roughly 17 kpc at the distance of M74.

In order to determine the extinction due to material
within M74 itself,
one might use SN 2003gd as a probe.
Both it and SN 2013ej exploded within the outer southern
arm of M74,
the former roughly 40 degrees farther along the arm 
from the center of the galaxy.
The similarity of the colors of these events suggests that they suffered
equally from reddening.
\citet{Hend2005}
use the colors of SN 2003gd itself, nearby stars, and
nearby HII regions to derive 
$E(B-V) = 0.14 \pm 0.06$;
this implies that the reddening contributions
from M74 and the Milky Way are roughly equal.

A more direct approach is to use high-resolution
spectra of SN 2013ej itself to measure the absorption
lines of neutral sodium (Na I), 
which are correlated with extinction along the 
line of sight.
\citet{Vale2014}
provide in their Figure 3 
a detailed graph of the spectrum centered on 
the NaI D lines.
As they state, this spectrum shows clearly 
the absorption lines due to gas within the Milky Way,
but no evidence for any absorption by gas in M74.
Using a digitized version of their spectrum,
I measure the equivalent widths of the Milky Way
components to be 
$\rm{EW(NaI\ D_1)} = 0.20 \AA$
and
$\rm{EW(NaI\ D_2)} = 0.26 \AA$.
The relationship in equation 9 of 
\citet{Pozn2012}
then yields 
$E(B-V) = 0.049 \pm 0.010.$
I will adopt this value for all following analysis.

Taking the relationships between reddening and extinction
given in 
\citet{Schl1998},
one can compute the extinction in each passband to be
$A_B = 0.20 \pm 0.04$,
$A_V = 0.15 \pm 0.03$,
$A_R = 0.12 \pm 0.02$,
and
$A_I = 0.08 \pm 0.02$.
If one were to choose the slightly higher reddening
given by 
\citet{Schl2011}
of $E(B-V) = 0.062$,
one would derive slightly larger extinctions of
$A_B = 0.27 \pm 0.05$,
$A_V = 0.21 \pm 0.04$,
$A_R = 0.17 \pm 0.03$,
and
$A_I = 0.12 \pm 0.03$.

Note that the adopted reddening is roughly $0.09$ mag
smaller than that of SN 2003gd,
which is consistent with the difference in
the $(B-V)$ colors of the two supernovae
during the plateau phase of their evolution.
Both the colors of the SN 2013ej and the high-resolution
spectra of 
\citet{Vale2014}
indicate that there was very little material along
the line of sight within M74,
and little circumstellar material surrounding
the progenitor itself.

\section{The distance to M74 and absolute magnitudes of SN 2013ej}

In order to calculate the absolute magnitude of SN 2013ej,
one must know the distance to its host galaxy.
Many attempts have been made to determine this distance,
using a variety of methods.
The appearance of the brightest individual stars 
has been used to derive distance moduli of
$(m - M) = 29.3$ \citep{Sohn1996},
$29.32$ \citep{Shar1996},
and
$29.44$ \citep{Hend2005}.
\citet{Sand1976} measured the angular sizes of the three largest
HII regions to estimate $(m - M) = 31.46$.
\citet{Hend2005}
applied the Standardised Candle Method of 
\citet{Hamu2002}
to spectra and photometry of SN 2003gd to 
derive 
$(m - M) = 29.9^{+0.6}_{-0.7}$;
they also determined a distance by assuming that
SNe 2003gd and 1999em were identical,
yielding
$(m - M) = 30.12 \pm 0.32$.
More recently,
\citet{Herr2008}
used the 
Planetary Nebula Luminosity Function (PNLF) to determine
a precise value of
$(m - M) = 29.67^{+0.06}_{-0.07}$.
\citet{Jang2014} kindly provided results in advance of
their publication of a distance based on the 
Tip of the Red Giant Branch (TRGB);
using HST images, they find
$(m - M) = 29.91 \pm 0.04\thinspace(\rm{rand}) \thinspace \pm 0.12\thinspace(sys)$.

\subsection{Applying the Expanding Photosphere Method (EPM) to SN 2013ej}

The Expanding Photosphere Method (EPM) applies basic
physics to determine the distance to a supernova
\citep{Kirs1974,Schm1992}.
Using spectra or photometry, one estimates the 
temperature of the photosphere at a set of times;
assuming that it radiates approximately as a blackbody,
one can compute the luminosity per unit area.
If the photosphere expands freely, then a combination
of radial velocity measurements and the time since
explosion permits one to compute the size of the 
photosphere.
One can multiply these quantities to determine the
luminosity of the photosphere,
then compare to the observed brightness to find
the distance to the event.

\begin{figure}
 \plotone{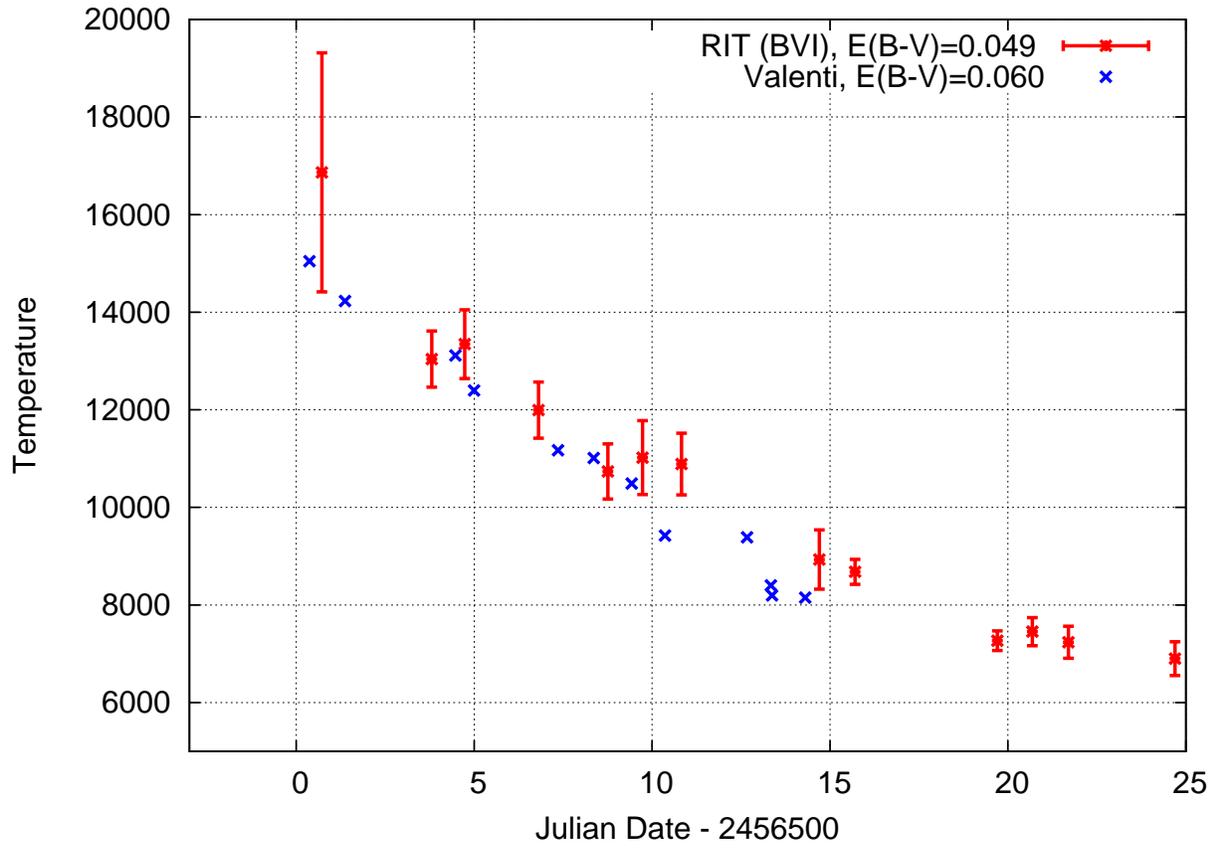}
 \caption{Temperature of SN 2013ej based on blackbody fits
          to $BVI$ photometry from RIT,
          and based on $UBgVrRiI$ photometry from 
          \citet{Vale2014}.
          \label{fig:temps} }
\end{figure}

Following the procedures described by
\citet{Bose2014},
I applied this technique to SN 2013ej.
The temperature was calculated based on $BVI$
photometry;
the $R$-band values were ignored, due to the presence
of strong H$\alpha$ features.
To estimate the uncertainties in the temperatures,
I used a Monte Carlo approach:
I generated thousands of instances of artificial
photometric measurements by adding random gaussian
noise to the actual magnitudes,
then fit blackbody spectra to those artificial
measurements.
The temperatures derived from RIT photometry
(after corrections for extinction)
are shown in 
Figure \ref{fig:temps};
they are slightly larger than those computed by
\citet{Vale2014},
which is somewhat surprising,
since my adopted reddening is smaller than that of 
\citet{Vale2014}.
However, both sets of temperatures, for
the most part, 
do agree within the uncertainties of the RIT values.
Since the RIT dataset lacks spectroscopy,
I adopted the radial velocities described in
\citet{Vale2014},
covering epochs 
$5 < JD - 2456500 < 22$.

\begin{figure}
 \plotone{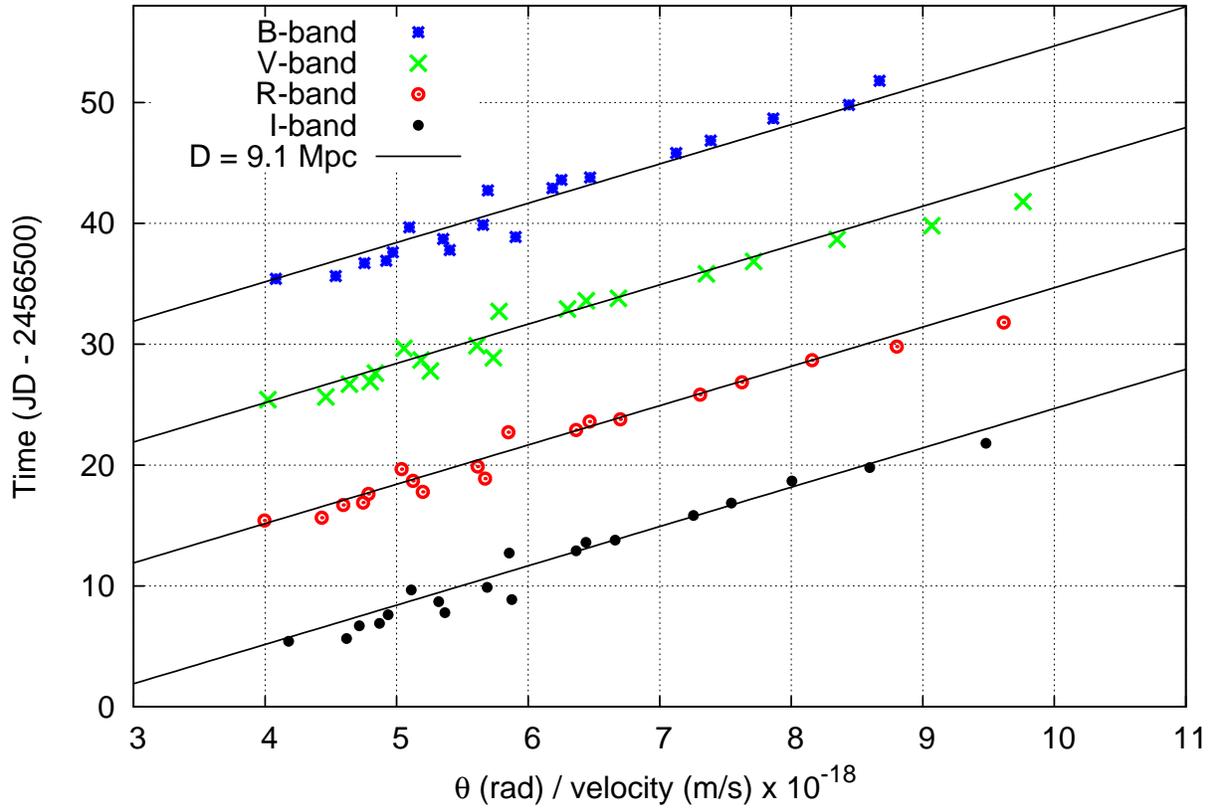}
 \caption{Distance to SN 2013ej based on EPM.
          The data have been shifted vertically
          for clarity by 30, 20, 10 days in 
          $B$, $V$, $R$, respectively.
          A line corresponding to the average distance
          $D = 9.1$ Mpc has been drawn to guide the eye.
          \label{fig:distances} }
\end{figure}

The procedures of 
\citet{Bose2014}
yield a semi-independent distance for each
passband of photometric measurements;
they are not fully independent due to the
photometric color corrections,
and due to the combination of magnitudes into
colors which are used to determine the temperature.
Plotting the time of each measurement against
the ratio of angular size to photospheric velocity
yields a graph in which the slope is the distance
to the supernova, and the y-intercept is the time
at which the size would be zero;
the actual time of explosion will be somewhat
later, since the star's initial size will always be
larger than zero.
Figure \ref{fig:distances}
shows the results of the analysis for all four passbands
of RIT photometry,
and Table \ref{tab:distances}
lists them.

\begin{center}
 \begin{deluxetable}{l r r}
   \tablecaption{Results of EPM applied to SN 2013ej \label{tab:distances} }
   \tablehead{
     \colhead{Passband} &
     \colhead{Distance (Mpc)} &
     \colhead{Time of explosion\tablenotemark{a} } 
   }

   \startdata

    B         &   $10.4 \pm 1.1$  &  -9.6  \\
    V         &   $8.5  \pm 0.8$  &  -6.4  \\
    R         &   $8.8  \pm 0.7$  &  -6.8  \\
    I         &   $9.4  \pm 0.9$  &  -8.2  \\

    \enddata
    \tablenotetext{a}{JD - 2456500; does not account for initial radius}
 \end{deluxetable}
\end{center}

The weighted average of these distances
is 
$D = 9.1 \pm 0.4$ Mpc,
corresponding to a distance modulus
$(m - M) = 29.79 \pm 0.11$.
One might conclude that the time of the
explosion is roughly 
$t_0 \sim 2456493$,
if one ignores the initial radius of the progenitor.
The rise time, from explosion to maximum light,
would then range from 14 days in $B$ to 26 days in $I$,
increasing monotonically with wavelength.
This is considerably shorter than the values
estimated for most of the sparsely-sampled
type IIP SNe modelled by
\citet{Sand2014},
but similar to the rise times for
the well-observed type IIP SN 2012aw
\citep{Bose2013}.

\subsection{Summary of distance measurements}

I give greatest weight to the PNLF 
\citep{Herr2008}
and TRGB
\citep{Jang2014}
methods,
and so adopt a distance modulus of
$(m - M) = 29.8 \pm 0.2$.
Using this value, and the extinction in each passband,
one can calculate the absolute magnitude of SN 2013ej 
at maximum light;
the results are shown in
Table \ref{tab:absmax}.

\begin{center}
 \begin{deluxetable}{l r }
   \tablecaption{Absolute magnitudes at maximum light, corrected for extinction \label{tab:absmax} }
   \tablehead{
     \colhead{Passband} &
     \colhead{mag\tablenotemark{a} } 
   }

   \startdata

    B         &   $-17.36 \pm 0.04  \pm 0.20 $ \\
    V         &   $-17.47 \pm 0.04  \pm 0.20 $ \\
    R         &   $-17.64 \pm 0.02  \pm 0.20 $ \\
    I         &   $-17.71 \pm 0.03  \pm 0.20 $ \\

    \enddata
    \tablenotetext{a}{absolute magnitude followed by random uncertainty, then systematic uncertainty}
 \end{deluxetable}
\end{center}

How does this event compare to other type IIP SNe?
\citet{Rich2002} examine the absolute magnitudes 
of 29 type IIP events,
finding a mean value $M_B = -17.00 \pm 1.12$.
It appears that SN 2013ej falls close to the
middle of this distribution,
indicating that it was typical of its class.

\section{Visual vs. CCD measurements}

Because SN 2013ej was one of the closest supernovae in 
the past few decades, 
it was monitored intensively by visual observers.
It provides us with a rare opportunity to compare 
visual measurements of a type IIP supernova to CCD $V$-band measurements.

\begin{figure}
 \plotone{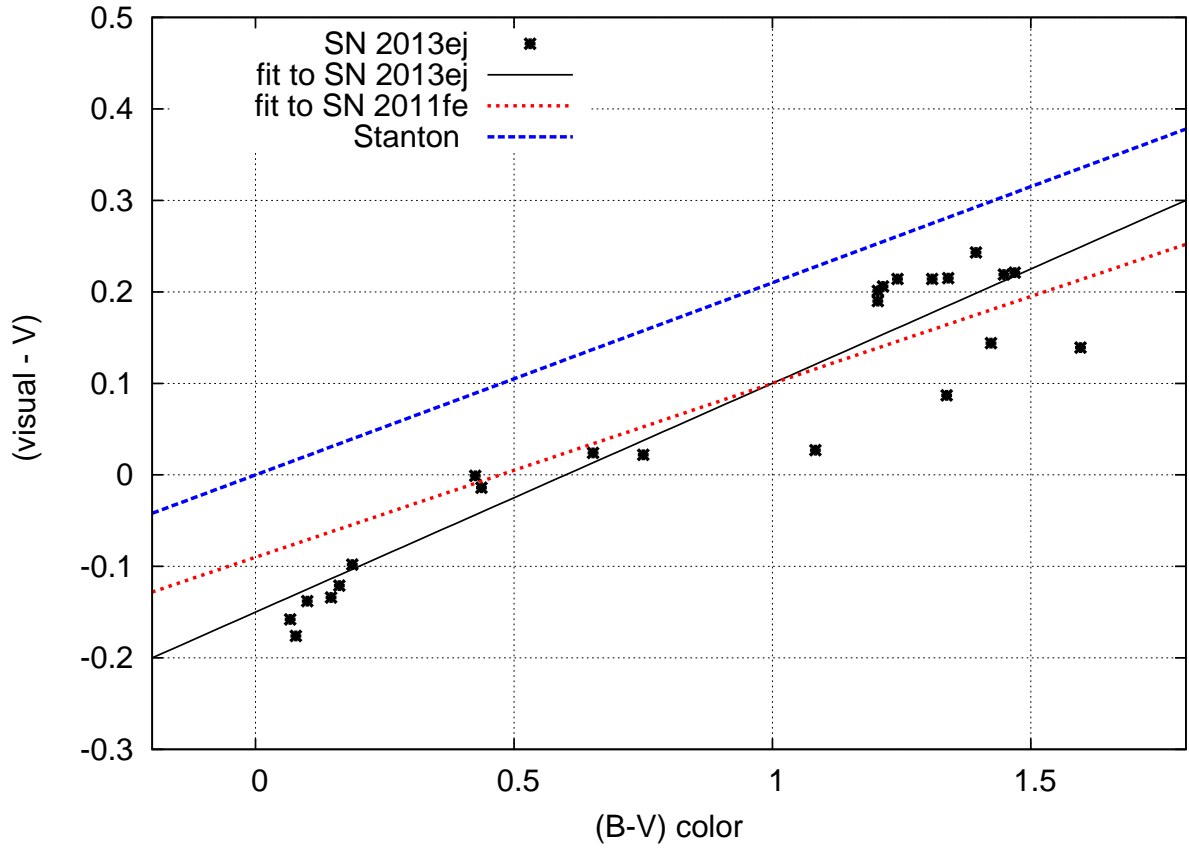}
 \caption{Difference between visual and CCD $V$-band
          measurements of SN 2013ej,
          together with relationships for SN 2011fe 
          \citep{Rich2012} and variable stars \citep{Stan1999}.
          \label{fig:visual} }
\end{figure}

I collected visual estimates from the AAVSO's
website \citep{Hend2014}.
There were a total of 119 measurements,
all with validation flag value 'Z', 
indicating that they had been checked only
for typos and data input errors.
The visual measurements cover the period
$ 1 < {\rm JD} - 2456500 < 105$,
which starts shortly before maximum light
and continues to the end of the plateau phase.
For each of the CCD V-band measurements,
I estimated a simultaneous visual magnitude
by fitting an unweighted low-order polynomial to the visual
measurements within $N$ days;
due to the decreasing frequency of visual measurements
and the less sharply changing light curve at late times,
the value $N$ was increased from 5 days to 8 days
at JD $2456540$ and again to 30 days at
JD $2456565$.
The differences between the polynomial
and each V-band measurement are shown
as a function of CCD $(B-V)$ color in
Figure \ref{fig:visual}.

An unweighted linear fit to these differences
yields the relationship
\begin{equation}
  ({\rm visual - V})_{\rm 2013ej} = -0.15 + \thinspace (0.25 \pm 0.02) * (B - V). \\
\end{equation}
This is very similar to the relationship between visual and CCD $V$-band
measurements of the type Ia SN 2011fe found by \citet{Rich2012}:
\begin{equation}
  ({\rm visual - V})_{\rm 2011fe} = -0.09 + \thinspace (0.19 \pm 0.04) * (B - V). \\
\end{equation}

The fact that two SNe of different type 
are perceived by human eyes in a similar fashion 
is consistent with the fact that
their light is dominated
by the continuum at these relatively early times.
In fact, the degree to which
eyes judge a supernova to be
fainter as it grows redder
agrees with the relationship for ordinary stars
measured by 
\citet{Stan1999},
further suggesting that human eyes
are responding primarily to the continuum emission
of supernovae.

\section{Conclusion}

Photometric $BVRI$ measurements 
from the RIT Observatory of SN 2013ej 
for six months
after its discovery show that it was a typical
type IIP supernova.
After correcting for extinction and assuming 
a distance modulus $(m - M) = 29.8$,
I find
absolute magnitudes of
$M_B = -17.36$,
$M_V = -17.47$,
$M_R = -17.64$,
and
$M_I = -17.71$.
Applying the expanding photosphere method
to this event yields a distance modulus
of 
$(m - M) = 29.79 \pm 0.11$,
agreeing well with other recent values.
The very low extinction along the line of sight,
and the proximity of its host galaxy M74,
make this one of the brightest core-collapse
supernovae since 1993.
As a result, many visual observers were able to 
monitor SN 2013ej for over three months;
the differences between their estimates and
CCD $V$-band measurements reveal the same 
trend with color that one sees in type Ia supernovae 
and in ordinary stars.

\acknowledgements
We thank Arne Henden and the staff at AAVSO for 
providing a sequence of comparison
stars near M74,
and the many observers who contribute their time,
energy, and measurements
to the AAVSO.
Stefano Valenti very gently pointed out an error in
the early version of this work and kindly provided
temperatures based on his own photometry.
Insung Jang cheerfully volunteered to choose
M74 as the next target in his project to measure
distances to nearby galaxies.
The anonymous referee made several suggestions
which improved this paper.
MWR is grateful for the continued support of the
RIT Observatory by RIT and its College of Science.

{\it Facilities:} \facility{AAVSO}, \facility{HST (ACS)}

\newpage

\end{document}